\documentclass[preprint]{elsarticle}
\usepackage[textwidth=19cm, textheight=24cm]{geometry}

\usepackage[utf8]{inputenc}
\usepackage[english]{babel}
\usepackage[T1]{fontenc}
\usepackage[hidelinks]{hyperref}
\usepackage{amsmath}
\usepackage{amssymb}
\usepackage{subcaption}
\usepackage{bm}
\usepackage[numbers]{natbib}
\usepackage{etoolbox}
\usepackage{siunitx}
\usepackage{algorithm}
\usepackage{algorithmicx}
\usepackage{algpseudocode}
\usepackage{setspace}
\usepackage{cancel}
\usepackage{nicefrac}
\usepackage{comment}
\usepackage{booktabs}

\hypersetup{pdfauthor=author}

\def\etal.{et\penalty50\ al.}

\makeatletter
\def\ps@pprintTitle{%
  \let\@oddhead\@empty
  \let\@evenhead\@empty
  \let\@oddfoot\@empty
  \let\@evenfoot\@oddfoot
}
\makeatother

\begin{document}

\newcommand{\mtm}{\textit{mode-to-mode} }
\newcommand{\mts}{\textit{mode-to-shell} }
\newcommand{\sts}{\textit{shell-to-shell} }
\renewcommand{\k}{\bm{\kappa}}
\newcommand{\kP}{\bm{\kappa'}}
\newcommand{\kmkP}{\bm{\kappa} - \bm{\kappa'}}
\def\Rey{\mbox{\textit{Re}}}
\newcommand{\volumeAve}[1]{\left\langle #1 \right\rangle_{\mathcal{L}}}

\begin{frontmatter}

\title{Viscous spectral energy coupling across scales in generalised Newtonian fluids}

\author[ifd,empa]{Arthur Couteau\corref{corauthor}}
\cortext[corauthor]{Corresponding author: \url{acouteau@ethz.ch}}

\author[empa]{Panayotis Dimopoulos Eggenschwiler}

\author[ifd]{Patrick Jenny}

\affiliation[ifd]{%
    organization={Institute of Fluid Dynamics, ETH Zürich},%
    city={Zürich},%
    postcode={CH-8092},%
    country={Switzerland}%
}

\affiliation[empa]{%
    organization={Chemical Energy Carriers and Vehicle Systems Laboratory, Empa},%
    city={Dübendorf},%
    postcode={CH-8600},%
    country={Switzerland}%
}

\begin{abstract}
We investigate the spectral energy dynamics of turbulent flows with variable viscosity using direct numerical simulation of homogeneous isotropic turbulence of generalised Newtonian fluids described by the Carreau constitutive model, covering both shear-thinning and shear-thickening regimes. The spectral evolution equations for the variable viscosity Navier-Stokes system show that the viscous term becomes nonlinear and gives rise to a convolution product in spectral space, formally analogous to that of the convective term. Unlike the constant viscosity case, where it acts as a purely local dissipation mechanism, the variable viscosity term carries both conservative (transfer) and non-conservative (dissipation) contributions entangled in the convolution product. We present novel computations of the viscous \mtm coupling $\hat{V}(\k, \kP)$, which does not satisfy a detailed conservation property analogous to that of the convective term. The viscous coupling maps reveal two distinct spectral regions: a sign-definite non-conservative region near $\kP \approx \bm{0}$, and a transfer-like dipole near $\kP \approx \k$ in shear-thickening fluids. The dipole satisfies the approximate antisymmetry $\hat{V}(\k, \kP) \approx -\hat{V}(\kP, \k)$, which is the defining signature of a conservative energy transfer. This demonstrates that energy transfer across scales, a role traditionally attributed exclusively to the convective nonlinearity, can arise from any nonlinear term in the momentum equation. The viscous energy transfer participates in the forward cascade alongside the convective transfer, eventually taking over the latter in the dissipation range. Its presence is connected to the emergence of power-law spectral decay replacing the classical exponential cutoff in shear-thickening fluids.
\end{abstract}


\end{frontmatter}

\section{Introduction}\label{sec:introduction}

Turbulent flows of generalised Newtonian fluids are encountered in a wide range of industrial and geophysical contexts, from polymer solutions and biological fluids to suspensions and magmatic flows. In such flows, the fluid viscosity is not a constant but a spatially and temporally varying field, coupled to the local rate of strain. This variable viscosity introduces an additional nonlinear term in the Navier-Stokes equations, a viscous coupling between modes, whose spectral structure and physical interpretation have received little attention.

In classical Newtonian turbulence, the spectral energy budget is governed by the interplay between two terms: the nonlinear convective term, which redistributes energy across scales through triadic interactions, and the linear viscous term, which dissipates energy locally at a rate proportional to $\nu \kappa^{2} \hat{E}(\kappa)$. The former is conservative, it satisfies a detailed conservation property ensuring that the energy lost by one mode is exactly gained by another \citep{onsagerStatisticalHydrodynamics1949,eyinkOnsagerTheoryHydrodynamic2006,lesieurTurbulenceFluids2008}, while the latter is purely dissipative and acts independently on each mode. This separation between conservative energy transfer and local dissipation underpins the Kolmogorov cascade picture \citep{KolmogorovLocalStructureTurbulence1941} and its many extensions \citep{frischTurbulenceLegacyKolmogorov1995,pope2000turbulent}. The distribution and locality of the convective energy transfer in spectral space have been the subject of extensive study, from the foundational work of \citet{kraichnanInertialrangeTransferTwo1971} and \citet{domaradzkiLocalEnergyTransfer1990,domaradzkiNonlocalTriadInteractions1992} to more recent direct computations of the \mtm energy transfer rate $\hat{T}(\k, \kP)$ by \citet{couteauInfluenceEnergycontainingScales2026}, who showed that the intensity of spectral energy transfers is governed by the energy content of the interacting modes, modulated by geometric constraints arising from the divergence-free condition in addition to the nonlocal nature of the interaction.

When the viscosity field is spatially variable, the mode-local dissipation is no more. The viscous stress tensor involves a product of the viscosity and strain rate fields, which in spectral space becomes a convolution. As a result, modes interact through the viscous term in a manner formally analogous to convective triad interactions, giving rise to spectral coupling terms that link energy non-locally across scales. Whether these couplings carry conservative effects, i.e., whether they can redistribute energy among modes without net production or destruction, is a non-trivial question with direct consequences for the spectral energy budget and its physical interpretation. To date, the viscous \mtm coupling $\hat{V}(\k, \kP)$ has never been computed or analysed.

The turbulence of non-Newtonian fluids has been studied primarily in wall-bounded configurations. Direct numerical simulations of pipe and channel flows of generalised Newtonian fluids have established that shear-thinning suppresses near-wall turbulence structures and reduces drag, while moderate shear-thickening has the opposite effect \citep{rudmanDirectNumericalSimulation2006,arosemenaTurbulentChannelFlow2021,arosemenaEffectsShearthinningRheology2021}. These studies focus on wall-normal statistics and do not address the spectral structure of the viscous term. In the context of viscoelastic turbulence, \citet{valenteEffectViscoelasticityTurbulent2014} and \citet{thaisSpectralAnalysisTurbulent2013} have identified polymer-induced modifications of the spectral energy budget, including an ``elasto-inertial'' cascade. However, the elastic stress introduces a separate energy reservoir with its own conservation equation, making the physics qualitatively different from the purely inelastic, strain-rate-dependent case considered here.

Recently, \citet{rostiEffectShearthinningScalings2025} studied homogeneous isotropic turbulence of shear-thinning Carreau fluids by DNS, demonstrating that Kolmogorov theory can be extended to such fluids provided that the mean viscosity is used to define the Kolmogorov scales. They showed that the $\kappa^{-5/3}$ inertial range is preserved and that the nonlinear energy flux is strengthened, while the dissipation range shrinks and small-scale universality is lost. Their study is restricted to shear-thinning fluids and does not examine the spectral structure of the viscous term itself.

In this work, we investigate the spectral energy rates of change arising from a variable viscosity field in homogeneous isotropic turbulence using DNS of generalised Newtonian fluids modelled by the Carreau constitutive law, covering both shear-thinning ($n < 1$) and shear-thickening ($n > 1$) regimes. We present novel computations of the viscous \mtm coupling $\hat{V}(\k, \kP)$, following the methodology developed by \citet{couteauInfluenceEnergycontainingScales2026} for the convective \mtm energy transfer. We derive the spectral evolution equations for the variable viscosity case and discuss the entanglement of conservative and non-conservative contributions in the viscous spectral coupling. The central finding is that the nonlinear viscous term can transfer energy across scales, a role traditionally attributed exclusively to the convective nonlinearity. More broadly, this demonstrates that any nonlinear term in the momentum equation, not only the convective one, can in principle produce inter-scale energy transfer through its spectral convolution product. We present turbulence statistics establishing that both rheological regimes follow a Kolmogorov-like cascading process upon rescaling with the mean viscosity, extending to shear-thickening fluids the results of \citet{rostiEffectShearthinningScalings2025}. While shear-thinning fluids retain a classical exponential decay in the dissipation range, we observe the emergence of power-law $\kappa^{-\alpha}$ scaling in the dissipation range of shear-thickening fluids, connected to the presence of viscous energy transfer in this spectral range.

\section{Governing equations and spectral formulation}\label{sec:formulation}

\subsection{Spectral evolution equations for variable viscosity flows}\label{sec:NS}

The Navier-Stokes momentum equation for an incompressible flow with variable viscosity reads
\begin{equation}
    \partial_{t} u_{j} + u_{l} \frac{\partial u_{j}}{\partial x_{l}} = - \frac{\partial p}{\partial x_{j}} + \frac{\partial}{\partial x_{l}} \left( 2 \nu S_{jl} \right),
    \label{eq:NS}
\end{equation}
with the strain rate tensor $S_{jl} = \tfrac{1}{2} (\partial u_{j}/\partial x_{l} + \partial u_{l}/\partial x_{j})$ and $\nu = \nu(\bm{x}, t)$ a spatially and temporally varying viscosity field. Using the Fourier series representation $u_{j}(\bm{x}, t) = \sum_{\k} \hat{u}_{j}(\k, t) \, e^{i \k \cdot \bm{x}}$ and $\nu(\bm{x}, t) = \sum_{\k} \hat{\nu}(\k, t) \, e^{i \k \cdot \bm{x}}$, the momentum equation becomes
\begin{equation}
    \partial_{t} \hat{u}_{j}(\k) = - P_{jm}(\k) \left[ \hat{G}_{m}^{\mathrm{conv}}(\k) - \hat{G}_{m}^{\mathrm{visc}}(\k) \right],
    \label{eq:NSspectral}
\end{equation}
with the Leray projection tensor $P_{jm}(\k) = \delta_{jm} - \kappa_{j} \kappa_{m}/\kappa^{2}$. The Fourier transform of the nonlinear convection term reads
\begin{equation}
    \hat{G}_{m}^{\mathrm{conv}}(\k) = i \kappa_{l} \sum_{\kP} \hat{u}_{m}(\kP) \hat{u}_{l}(\k - \kP).
    \label{eq:convterm}
\end{equation}
Since the viscosity is a spatially variable scalar field, the viscous term is also nonlinear. Its Fourier transform reads
\begin{equation}
    \hat{G}_{m}^{\mathrm{visc}}(\k) = 2 i \kappa_{l} \sum_{\kP} \hat{\nu}(\kP) \hat{S}_{ml}(\kmkP).
    \label{eq:viscterm}
\end{equation}
This is a convolution product, coupling all modes in the domain through the viscosity field. Using the spectral strain rate tensor $2\hat{S}_{ml}(\k) = i(\kappa_{m} \hat{u}_{l}(\k) + \kappa_{l} \hat{u}_{m}(\k))$, the viscous term is expanded as
\begin{equation}
    \hat{G}_{m}^{\mathrm{visc}}(\k) = - \kappa_{l} \sum_{\kP} \hat{\nu}(\kP) \left[ (\kmkP)_{m} \hat{u}_{l}(\kmkP) + (\kmkP)_{l} \hat{u}_{m}(\kmkP) \right].
    \label{eq:viscterm_expanded}
\end{equation}
For constant viscosity, the only non-zero contribution of the convolution sum is for $\kP = \bm{0}$, and one recovers the classical spectral diffusion term $\hat{G}_{m}^{\mathrm{visc}}(\k) = -\hat{\nu}(\bm{0}) \kappa^{2} \hat{u}_{m}(\k)$ with the help of the divergence-free condition $\kappa_{j} \hat{u}_{j}(\k) = 0$.

The kinetic energy of mode $\k$ is $\hat{E}(\k) = \tfrac{1}{2} \hat{u}_{j}(\k) \hat{u}_{j}^{*}(\k)$. Its evolution follows from (\ref{eq:NSspectral}) as
\begin{equation}
    \partial_{t} \hat{E}(\k) = - \sum_{\kP} \hat{T}(\k, \kP) + \sum_{\kP} \hat{V}(\k, \kP),
    \label{eq:modalEnergyConservation}
\end{equation}
where the convective and viscous \mtm energy rates of change are given by
\begin{equation}
    \hat{T}(\k, \kP) = \Re\left\{ i \kappa_{l}\, \hat{u}_{l}(\kmkP)\, \hat{u}_{j}(\kP)\, \hat{u}_{j}^{*}(\k) \right\},
    \label{eq:Tdef}
\end{equation}
\begin{equation}
    \hat{V}(\k, \kP) = \Re\left\{ 2 i \kappa_{l}\, \hat{\nu}(\kP)\, \hat{S}_{jl}(\kmkP)\, \hat{u}_{j}^{*}(\k) \right\}.
    \label{eq:Vdef}
\end{equation}

\subsection{Physical interpretation: transport, transfer and coupling}\label{sec:interpretation}

We must carefully distinguish the physical meaning of $\hat{T}(\k, \kP)$ and $\hat{V}(\k, \kP)$. In what follows, we use ``transport'' to refer to the movement of kinetic energy through physical space, ``transfer'' to refer to the movement of kinetic energy between spectral modes. Both are conservative phenomena. We use ``coupling'' to refer to the general triad interaction that may contain both conservative and non-conservative effects.

\begin{figure}
    \centering
    \includegraphics[]{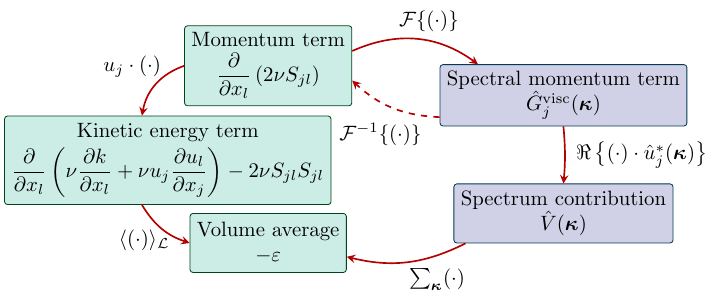}
    \caption{Relations between the physical and spectral quantities for the nonlinear viscous term. There is, in general, no direct path from a single physical kinetic energy term to its spectrum, as the latter is obtained from the transformation of a momentum term.}
    \label{fig:transformVSSpectrumViscousVariable}
\end{figure}

A linear term in the physical momentum equation results, upon Fourier transformation, in a single local contribution to the spectral energy budget. A nonlinear term produces a convolution product, implying coupling between all modes in the domain. Separately, a conservative term (one that can be written in divergence form) results in a net zero energy change upon volume averaging, while a non-conservative term contributes to net production or dissipation. These last two definitions are exact in triply-periodic domains, as is the case in this study, but are also valid on average when considering statistically homogeneous turbulence.

In the constant viscosity case, these properties are clearly separated. The convective term is both nonlinear and conservative: it produces a convolution product while conserving energy through the detailed conservation property $\hat{T}(\k, \kP) = -\hat{T}(\kP, \k)$ \citep{onsagerStatisticalHydrodynamics1949,eyinkOnsagerTheoryHydrodynamic2006,couteauInfluenceEnergycontainingScales2026} ensures that the energy lost by mode $\k$ is exactly gained by mode $\kP$, such that $\sum_{\k} \sum_{\kP} \hat{T}(\k, \kP) = 0$. This property allows the unambiguous interpretation of $\hat{T}(\k, \kP)$ as a pure energy transfer rate between modes $\k$ and $\kP$. The viscous term, on the other hand, is linear and non-conservative: it produces a purely local, mode-independent dissipation $\hat{D}(\k) = 2 \nu \kappa^{2} \hat{E}(\k)$.

The introduction of a variable viscosity field changes the picture fundamentally. The viscous term becomes nonlinear, giving rise to the convolution product (\ref{eq:viscterm}), and remains non-conservative since the physical momentum term $\partial_{l}(2\nu S_{jl})$ result in a sum of pure divergence (transport) and sink (dissipation) terms in the conservation equation of kinetic energy (see figure \ref{fig:transformVSSpectrumViscousVariable}). As a result, the spectral quantity $\hat{V}(\k, \kP)$ inherits both effects, entangled in the convolution product. The viscous coupling therefore carries mixed effects of both conservative energy transfer across scales and non-conservative energy production or dissipation. This entanglement is highlighted by the absence of a detailed conservation property. Writing out the full expression,
\begin{equation}
    \hat{V}(\k, \kP) = \Re\left\{ - \kappa_{l}\, \hat{\nu}(\kP) \left[ (\kmkP)_{j}\, \hat{u}_{l}(\kmkP) + (\kmkP)_{l}\, \hat{u}_{j}(\kmkP) \right] \hat{u}_{j}^{*}(\k) \right\},
    \label{eq:Vfull}
\end{equation}

\begin{equation}
    \hat{V}(\kP, \k) = \Re\left\{ - \kappa'_{l}\, \hat{\nu}(\k) \left[ (\kP - \k)_{j}\, \hat{u}_{l}^{*}(\kmkP) + (\kP - \k)_{l}\, \hat{u}_{j}^{*}(\kmkP) \right] \hat{u}_{j}^{*}(\kP) \right\},
    \label{eq:Vsym}
\end{equation}
it is verified by direct inspection that $\hat{V}(\k, \kP) \neq -\hat{V}(\kP, \k)$ in general, since the roles of the viscosity and velocity modes are not interchangeable. We make the following ansatz
\begin{equation}
    \hat{V}(\k, \kP) = \hat{V}^{\mathrm{cons}}(\k, \kP) + \hat{V}^{\mathrm{n\text{-}cons}}(\k, \kP),
    \label{eq:Vdecomp}
\end{equation}
where no closed-form expression exists for either component separately. The spectral quantity $\hat{V}(\k, \kP)$ is obtained from the transformation of a physical momentum term into a spectral momentum term, further multiplied by $\hat{u}_{j}^{*}(\k)$ and projected onto the real axis. There exists no operation that allows one to go from a single, well-defined term in the physical kinetic energy equation directly to its spectrum. Since the nonlinear viscous term cannot be separated at the momentum equation level into a pure nonlinear transport term and a pure non-conservative term, the corresponding spectral quantity inherits both contributions.

This observation extends to any nonlinear term in the momentum equation. A nonlinear term always produces a convolution product in spectral space, and this convolution product will, in general, carry both conservative and non-conservative contributions. The convection coupling $\hat{T}(\k, \kP)$ is particular and by construction only has a conservative contribution. Whether the conservative part is significant, i.e., whether genuine inter-scale energy transfer occurs through the term, depends on the specific physics of the coupling and can only be assessed by direct computation of the \mtm coupling. The antisymmetry $\hat{X}(\k, \kP) = -\hat{X}(\kP, \k)$ is, by definition, the signature of a conservative energy transfer: it means that whatever energy is gained by mode $\k$ through interaction with $\kP$ is lost by $\kP$ through the corresponding interaction with $\k$. We will show that this antisymmetry emerges in the viscous coupling under certain conditions, specifically in shear-thickening fluids and in defined spectral regions, providing evidence that energy transfer across scales from viscous effects is a real physical phenomenon.

\subsection{Dissipation rate decomposition}\label{sec:dissipation}

The total dissipation rate is $\varepsilon = \volumeAve{2 \nu S_{jl} S_{jl}}$, where $\volumeAve{\cdot}$ indicates volume averaging. Since the viscosity field is positive definite, $\varepsilon$ remains positive and balances the energy produced by the forcing scheme in statistical equilibrium. We split the viscosity field as
\begin{equation}
    \nu = \volumeAve{\nu} + \nu',
\end{equation}
where $\volumeAve{\nu}$ is the volume-averaged viscosity and $\nu'$ the fluctuating part. The total dissipation rate then decomposes as
\begin{equation}
    \varepsilon = \varepsilon_{\volumeAve{\nu}} + \varepsilon_{\nu'},
\end{equation}
with
\begin{equation}
    \varepsilon_{\volumeAve{\nu}} = 2 \volumeAve{\nu} \volumeAve{S_{jl} S_{jl}}, \qquad
    \varepsilon_{\nu'} = \varepsilon - \varepsilon_{\volumeAve{\nu}}.
\end{equation}
Since $\volumeAve{\nu} > 0$, we have $\varepsilon_{\volumeAve{\nu}} \geq 0$. However, there is no constraint on the sign of $\varepsilon_{\nu'}$. The spectral counterparts of these decompositions are obtained by selecting contributions in the convolution sum. The mean-viscosity contribution corresponds to the $\kP = \bm{0}$ term,
\begin{equation}
    \hat{D}_{\volumeAve{\nu}}(\k) = \hat{V}(\k, \bm{0}) = 2 \volumeAve{\nu}\, \kappa^{2}\, \hat{E}(\k),
\end{equation}
which has the same form as the classical Newtonian dissipation spectrum but with $\volumeAve{\nu}$ replacing the constant viscosity. The fluctuating contribution collects all remaining terms,
\begin{equation}
    \hat{D}_{\nu'}(\k) = \sum_{\kP \neq \bm{0}} \hat{V}(\k, \kP).
    \label{eq:Dnuprime}
\end{equation}
The spectra of both contributions sum to the total dissipation,
\begin{equation}
    \varepsilon_{\volumeAve{\nu}} = \sum_{\k} \hat{D}_{\volumeAve{\nu}}(\k), \qquad
    \varepsilon_{\nu'} = \sum_{\k} \hat{D}_{\nu'}(\k).
\end{equation}
The fluctuating dissipation spectrum $\hat{D}_{\nu'}$ can therefore be understood as a measure of the deviation of the true dissipation spectrum from a classical Newtonian-like spectrum based on the mean viscosity.

\subsection{Carreau viscosity model}\label{sec:carreau}

The viscosity field must be specified by a constitutive model. In this work, we adopt the Carreau model \citep{carreauRheologicalEquationsMolecular1972}, an isotropic generalised Newtonian model that describes the viscosity as a function of the local rate of strain $\dot{\gamma} = \sqrt{2 S_{ij} S_{ij}}$,
\begin{equation}
    \nu = \nu_{\infty} + (\nu_{0} - \nu_{\infty}) \left[ 1 + (\mathcal{K} \dot{\gamma})^{2} \right]^{(n - 1)/2},
    \label{eq:Carreau}
\end{equation}
where $\nu_{0}$ and $\nu_{\infty}$ are the asymptotic viscosities at zero and infinite rate of strain, $\mathcal{K}$ is the consistency index, and $n$ is the power index. A Newtonian fluid is recovered for $n = 1$, shear-thinning for $n < 1$, and shear-thickening for $n > 1$. The Carreau model is preferred over the simpler power-law model $\nu = \mathcal{K} \dot{\gamma}^{n-1}$ because it provides finite viscosity values at zero and infinite rate of strain, avoiding numerical issues with singular behaviour at vanishing velocity gradients.

\subsection{Numerical method and simulation parameters}\label{sec:numerics}

The simulations are performed using a pseudo-spectral DNS code on a triply periodic cubic domain. The convective nonlinearity is computed using the standard pseudo-spectral method with phaseshift dealiasing. For the nonlinear viscous term, we adopt a single-pass approach: the strain rate tensor is computed in spectral space and transformed to physical space, where all nonlinear operations (computation of $\dot{\gamma}$, evaluation of the Carreau model, and multiplication by the strain rate) are performed before transformation back to spectral space and dealiasing by phase shifting. This approach ensures that physically positive quantities, such as $\dot{\gamma}^{2}$, remain positive throughout the computation, preserving the realisability of the solution.

The flow is sustained by a low-wavenumber random forcing scheme constructed from an Ornstein-Uhlenbeck process \citep{eswaranExaminationForcingDirect1988}, injecting energy in the shell $\kappa_{f} \in ]0, 2\sqrt{2}]$. The statistics are gathered once the flow has reached statistical stationarity. The viscous \mtm coupling $\hat{V}$ is computed directly from (\ref{eq:Vfull}) using the same algorithm as for the convective \mtm energy transfer $\hat{T}$ described in \citet{couteauInfluenceEnergycontainingScales2026}, and averaged over 1500 realisations spanning more than 20 large-eddy turnover times. The obtained $\hat{V}(\k,\kP)$ maps are validated by comparing their sum to the value obtained from the pseudo-spectral method. The \mts transfer functions, measuring the energy transfer rate between the mode $\k$ and a shell $K'$, are defined as
\begin{equation}
    \hat{M}^{\mathrm{conv}}(\k, K') = \sum_{\kP \in K'} \hat{T}(\k, \kP) \qquad \hat{M}^{\mathrm{visc}}(\k, K') = \sum_{\kP \in K'} \hat{V}(\k, \kP)
\end{equation}

\begin{table}
    \centering
    \caption{Simulation parameters for all cases. The resolution is $N^{3}$ grid points. $Re_{\lambda}$ is the Taylor-scale Reynolds number based on $\volumeAve{\nu}$.}
    \label{tab:simparams}
    \begin{tabular}{@{}lcccccc@{}}
    \toprule
    Case & $n$ & $\nu_{0}$ & $\nu_{\infty}$ & $\mathcal{K}$ & $N$ & $Re_{\lambda}$ \\
    \midrule
    TH04 & 0.4 & $5 \times 10^{-3}$ & $5 \times 10^{-5}$ & 1.0 & 512 & 170 \\
    TH07 & 0.7 & $5 \times 10^{-3}$ & $5 \times 10^{-5}$ & 1.0 & 512 & 130 \\
    N10   & 1.0 & $5 \times 10^{-3}$ & --- & --- & 512 & 110 \\
    TK15 & 1.5 & $5 \times 10^{-3}$ & --- & 1.0 & 512 & 70 \\
    TK30 & 3.0 & $5 \times 10^{-3}$ & --- & 1.0 & 512 & 35 \\
    TK30-LV1 & 3.0 & $1 \times 10^{-4}$ & --- & 1.0 & 512 & 70 \\
    TK30-LV2 & 3.0 & $2 \times 10^{-5}$ & --- & 1.0 & 512 & 105 \\
    \bottomrule
    \end{tabular}
\end{table}

Table \ref{tab:simparams} summarises the simulation parameters. The shear-thinning cases (TH04, TH07) have $n < 1$, such that the viscosity decreases in regions of high strain rate. The shear-thickening cases (TK15, TK30) have $n > 1$, producing the opposite behaviour. Two additional shear-thickening cases at lower base viscosity (TK30-LV1, TK30-LV2) are included to investigate Reynolds number effects on the dissipation range scaling.

\section{Turbulence statistics}\label{sec:statistics}

\subsection{Flow visualisations}\label{sec:visualisations}

\begin{figure}
    \centering
    \includegraphics[]{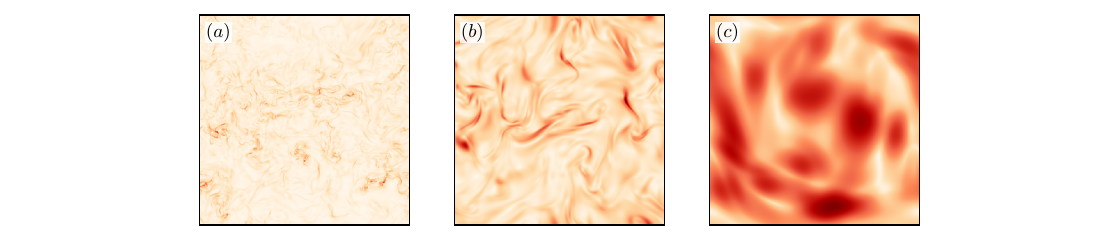}
    \caption{Instantaneous rate of strain fields for ($a$) shear-thinning ($n = 0.4$), ($b$) Newtonian ($n = 1.0$), and ($c$) shear-thickening ($n = 3.0$) fluids. The base viscosity is $\nu_{0} = 5\times10^{-3}$ for all cases.}
    \label{fig:rateOfStrain}
\end{figure}

Figure \ref{fig:rateOfStrain} shows instantaneous rate of strain fields for a shear-thinning, Newtonian, and shear-thickening fluid. In the shear-thinning case, the flow displays markedly finer structures than the Newtonian reference. The locally reduced viscosity in high-strain regions allows sharper velocity gradients to form, which in turn further reduce the local viscosity in a positive feedback loop that only terminates when dissipation ultimately prevails. The shear-thickening case exhibits the opposite trend: the flow appears smoother, with small-scale features suppressed and large-scale structures more prominent. The locally increased viscosity in high-strain regions acts to dampen the velocity gradients that produced it, creating a self-regulating mechanism in which fine-scale motions are more efficiently suppressed than in the Newtonian case. These substantial changes in coherent structures are highly relevant in many complex flows; for example, in turbulent suspensions, heavy particles cluster in flow regions of low vorticity and high strain \citep{maxeyGravitationalSettlingAerosol1987,squiresPreferentialConcentrationParticles1991}.

\subsection{Kinetic energy spectra}\label{sec:spectra}

Since all simulations are driven by a random forcing scheme using the same parameters and have reached statistical stationarity, the total dissipation rate $\varepsilon$ is, on average, the same across all cases. What differs substantially is the way in which the energy and the dissipation are distributed across scales.

\begin{figure}
    \centering
    \includegraphics[]{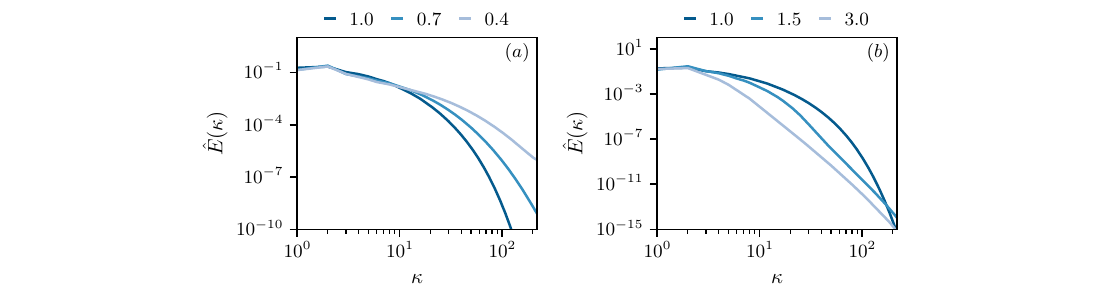}
    \caption{Kinetic energy spectra $\hat{E}(\kappa)$ for (a) shear-thinning and (b) shear-thickening fluids}
    \label{fig:kinEnergySpectra}
\end{figure}

Figure \ref{fig:kinEnergySpectra} shows the kinetic energy spectra for all five cases with $\nu_{0} = 5 \times 10^{-3}$. The separate kinetic energy spectra with the scaling ranges highlighted are found in App. \ref{sec:app:KineticEnergySpectra}. The Newtonian case displays the classical $\kappa^{-5/3}$ inertial range followed by a sharp exponential decay in the dissipation range. For the shear-thinning cases, we obtain results consistent with those of \citet{rostiEffectShearthinningScalings2025}: a systematic extension of the inertial range to higher wavenumbers as $n$ decreases, and a less steep decay at small scales. This is consistent with the visual observation of finer structures and with the reduced effective viscosity in high-strain regions.

The shear-thickening cases exhibit a qualitatively different behaviour. The $\kappa^{-5/3}$ inertial range shrinks with increasing $n$, and a new scaling regime emerges at high wavenumbers. We observe power-law behaviour $\hat{E}(\kappa) \sim \kappa^{-\alpha}$ in this dissipation scaling range, with fitted exponents $\alpha \approx 9$ for $n = 1.5$ and $\alpha \approx 8$ for $n = 3.0$. These exponents are obtained from fitting the numerical data and are not derived from any theoretical argument. An interesting and somewhat counter-intuitive observation is that the power-law decay, while steep, is slower than an exponential, so there is actually more energy at the very smallest scales in the shear-thickening case than in the Newtonian one. The apparent smoothness of the shear-thickening flow fields arises from a depletion of energy at intermediate wavenumbers, not at the smallest ones.

\subsection{Spectral energy budget}\label{sec:budget}

\begin{figure}
    \centering
    \includegraphics[]{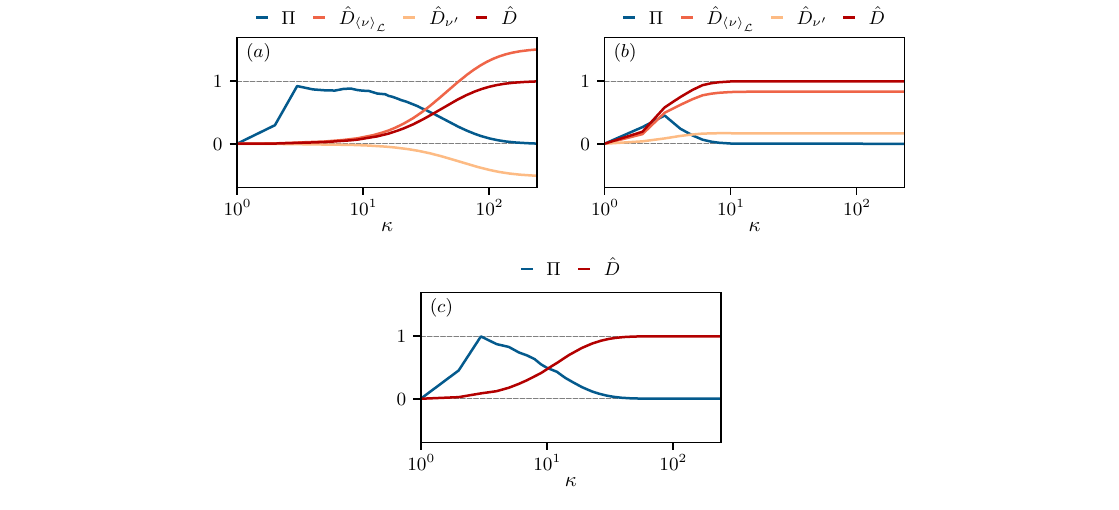}
    \caption{Kinetic energy spectral fluxes normalised by the total dissipation rate $\varepsilon$. The convective flux $\Pi$ (blue), and the cumulative dissipation contributions from the mean viscosity $\hat{D}_{\volumeAve{\nu}}$ (dark orange) and the viscosity fluctuations $\hat{D}_{\nu'}$ (light orange) are shown. The total dissipation $\hat{D}$ is in red.}
    \label{fig:spectralFluxes}
\end{figure}

Figure \ref{fig:spectralFluxes} presents the spectral fluxes normalised by the total dissipation rate $\varepsilon$ for representative shear-thinning, Newtonian, and shear-thickening cases. In the Newtonian case, the convective spectral flux $\Pi$ rises from the forcing scales, reaches a plateau of approximately unity in the inertial range (reflecting a constant energy transfer rate equal to $\varepsilon$), and the cumulative dissipation increases monotonically, as per the classical cascade picture.

In the shear-thinning cases, the viscosity fluctuation contribution $\hat{D}_{\nu'}$ takes negative values, acting as a production-like mechanism. An interpretation is that the mean viscosity $\volumeAve{\nu}$ overestimates the effective viscosity in high-strain regions where dissipation is most active: the mean-viscosity spectrum $\hat{D}_{\volumeAve{\nu}}$ therefore overpredicts the true dissipation, and $\hat{D}_{\nu'}$ provides a negative correction. This observation is consistent with the results of \citet{rostiEffectShearthinningScalings2025}. It is also consistent with the negative correlation between viscosity and velocity gradients in shear-thinning fluids: the physical cross terms in the kinetic energy equation involving viscosity gradients take negative values, and these sign-definite contributions are expected to be reflected in $\hat{D}_{\nu'}$. We note that similar production-like effects from viscosity fluctuations have been reported in wall-bounded flows of shear-thinning fluids \citep{arosemenaTurbulentChannelFlow2021}, where they modify the turbulent kinetic energy budget primarily through the mean-shear turbulent viscous transport.

In the shear-thickening cases, the inverse effect is observed. The local viscosity is enhanced in high-strain regions, so the mean viscosity underestimates the effective dissipation, and $\hat{D}_{\nu'}$ provides a positive correction. Both contributions are positive and together account for the total dissipation rate.

\subsection{Dual scaling range and Reynolds number effects}\label{sec:dualscaling}

\begin{figure}
    \centering
    \includegraphics[]{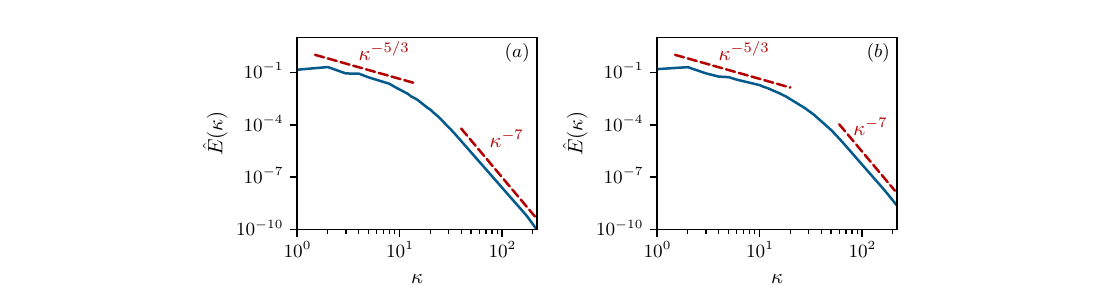}
    \caption{Kinetic energy spectra for shear-thickening Carreau fluid with $n = 3.0$ at lower base viscosities: $\nu_{0} = 1\times10^{-4}$ (left) and $\nu_{0} = 2\times10^{-5}$ (right).}
    \label{fig:lowerViscSpectra}
\end{figure}

\begin{figure}
    \centering
    \includegraphics[]{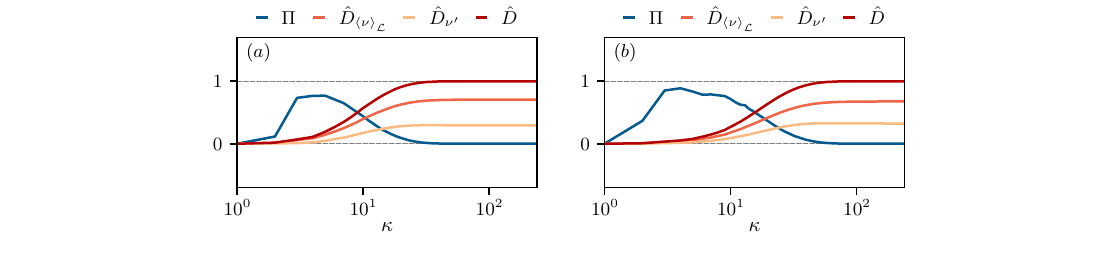}
    \caption{Spectral fluxes for shear-thickening Carreau fluid with $n = 3.0$ at lower base viscosities: $\nu_{0} = 1\times10^{-4}$ (left) and $\nu_{0} = 2\times10^{-5}$ (right).}
    \label{fig:lowerViscFluxes}
\end{figure}

The case TK30 ($n = 3.0$, $\nu_{0} = 5 \times 10^{-3}$) does not display a proper inertial range: the spectral flux $\Pi$ never reaches a constant plateau. To investigate whether this is a finite Reynolds number effect, we perform two additional simulations at $\nu_{0} = 1 \times 10^{-4}$ and $\nu_{0} = 2 \times 10^{-5}$ (figures \ref{fig:lowerViscSpectra} and \ref{fig:lowerViscFluxes}). As the base viscosity decreases, a clear $\kappa^{-5/3}$ inertial range emerges, accompanied by a well-defined plateau in $\Pi$, confirming that the absence of an inertial range at higher $\nu_{0}$ was a finite Reynolds number effect rather than a fundamental feature of shear-thickening turbulence. The dissipation scaling range persists past the inertial range, but the exponent is now closer to $\kappa^{-7}$ in both cases, compared to $\kappa^{-8}$ at the higher base viscosity. This suggests that $\alpha$ depends on the Reynolds number as well as on $n$, and may converge as the Reynolds number increases.

The spectral budget reveals that the contribution of $\hat{D}_{\nu'}$ to the total dissipation grows as the base viscosity decreases. This is expected: at lower base viscosities, more intense strain-rate fluctuations produce larger viscosity fluctuations through the shear-thickening constitutive law.

\subsection{Kolmogorov scaling collapse}\label{sec:collapse}

\begin{figure}
    \centering
    \includegraphics[]{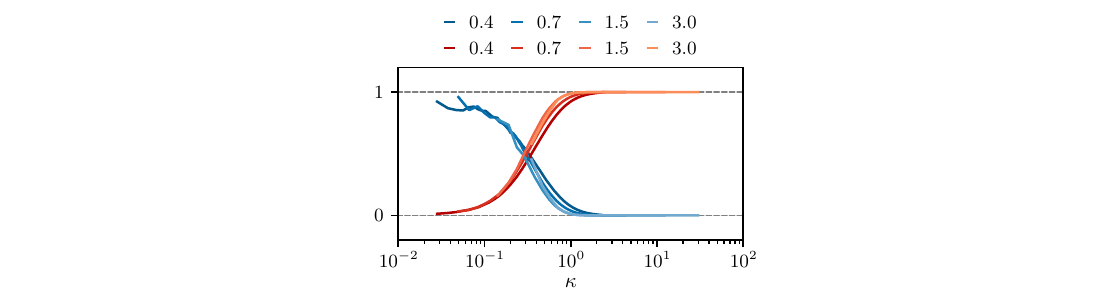}
    \caption{Kinetic energy spectral fluxes normalised by the effective Kolmogorov length $\eta_{\mathrm{eff}} = (\volumeAve{\nu}^{3}/\varepsilon)^{1/4}$. The convective flux $\Pi$ is shown in blue shades and the total dissipation $\hat{D}$ in red shades.}
    \label{fig:rescaledFluxes}
\end{figure}

Following \citet{rostiEffectShearthinningScalings2025}, the spectral fluxes can be normalised using an effective Kolmogorov length scale $\eta_{\mathrm{eff}} = (\volumeAve{\nu}^{3}/\varepsilon)^{1/4}$. Under this rescaling (figure \ref{fig:rescaledFluxes}), the crossover between the convective transport regime and the dissipative regime collapses onto a single location at $\kappa \eta_{\mathrm{eff}} \approx 0.3$ for all cases, consistent with classical Kolmogorov theory and with the value observed in Newtonian turbulence \citep{pope2000turbulent}. This collapse was previously reported for shear-thinning fluids by \citet{rostiEffectShearthinningScalings2025}; we show here that it extends to shear-thickening fluids as well. This indicates that, despite the substantial spatial variability of the viscosity field, the scale at which the inertial range gives way to the viscous range is still well predicted by the Kolmogorov scale based on the average viscosity.

\section{Viscous spectral coupling}\label{sec:coupling}

\subsection{Mode-to-mode coupling maps}\label{sec:m2mmaps}

\begin{figure}
    \centering
    \includegraphics[]{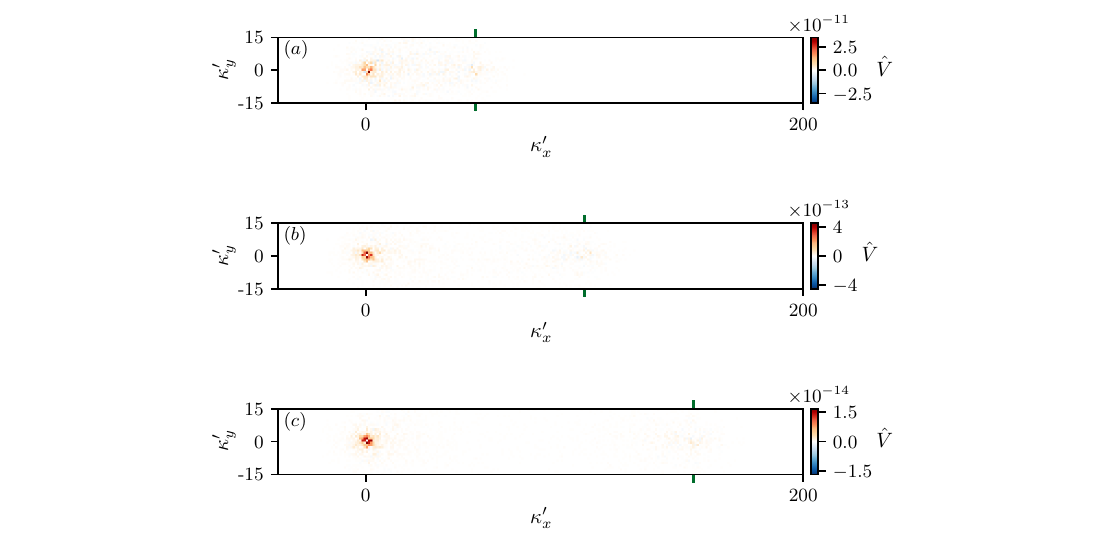}
    \caption{Viscous \mtm spectral coupling $\hat{V}(\k, \kP)$ for a shear-thinning fluid with $n = 0.7$. The green ticks indicate the sampling point $\k$.}
    \label{fig:M2M_TH07}
\end{figure}

\begin{figure}
    \centering
    \includegraphics[]{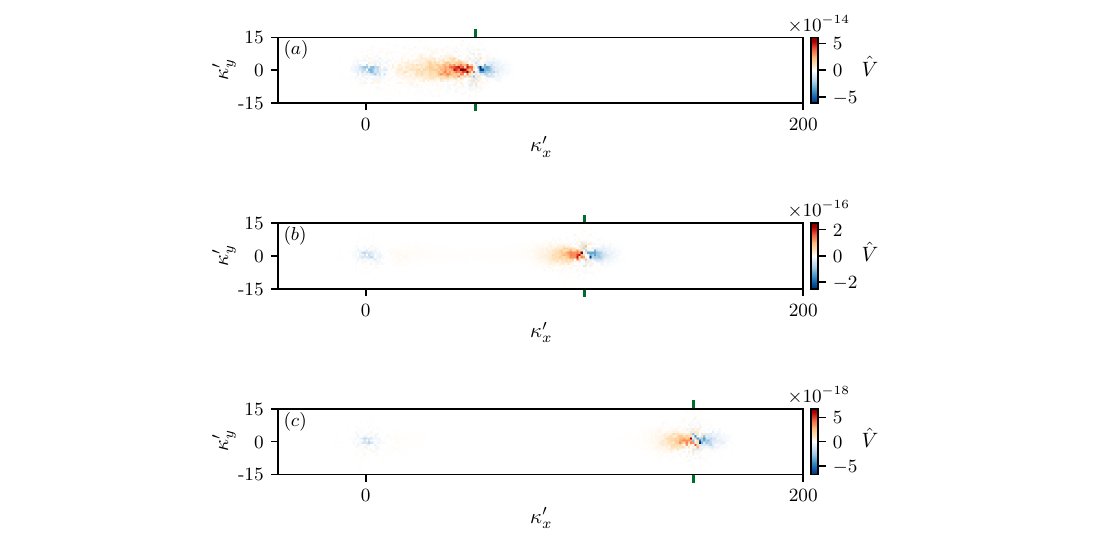}
    \caption{Viscous \mtm spectral coupling $\hat{V}(\k, \kP)$ for a shear-thickening fluid with $n = 1.5$. The green ticks indicate the sampling point $\k$.}
    \label{fig:M2M_TK15}
\end{figure}

Figures \ref{fig:M2M_TH07} and \ref{fig:M2M_TK15} present the viscous \mtm spectral coupling $\hat{V}(\k, \kP)$ for a shear-thinning ($n = 0.7$) and a shear-thickening ($n = 1.5$) fluid, for sampling wavenumbers $\k = [50, 0]$, $\k = [100, 0]$ and $\k = [150, 0]$. The point $\hat{V}(\k, \bm{0})$, corresponding to dissipation by the mean viscosity, is excluded from the visualisation as its magnitude far exceeds the remaining contributions. Two spectral regions of interest are identified.

\subsubsection{Non-conservative coupling near the origin.}
In all non-Newtonian cases, non-zero values of $\hat{V}(\k, \kP)$ appear around $\kP \approx \bm{0}$. The position of this region does not depend on the sampling wavenumber $\k$. By the structure of the convolution product (\ref{eq:Vdef}), contributions from this region correspond to viscous triad interactions involving low-wavenumber viscosity modes $\hat{\nu}(\kP)$ and strain-rate tensor modes $\hat{S}_{jl}(\kmkP)$ at wavenumbers close to that of the sampling point. This is a nonlocal coupling in which the large-scale structure of the viscosity field modulates the dissipation of individual modes. The smallest viscosity wavenumber is $\kP = \bm{0}$, which gives the purely non-conservative mean-viscosity dissipation $\hat{D}_{\volumeAve{\nu}}(\k) = 2 \volumeAve{\nu}\, \kappa^{2}\, \hat{E}(\k)$ and sits at the centre of this region.

The near-origin region is predominantly sign-definite. In the shear-thinning cases, the energy rates of change are positive, indicating a net production-like effect, consistent with the negative $\hat{D}_{\nu'}$ discussed in \S\ref{sec:budget}. In the shear-thickening cases, they are of opposite sign, consistent with the positive (dissipative) $\hat{D}_{\nu'}$. We therefore associate this region to the non-conservative contributions and state
\begin{equation}
    |\hat{V}^{\mathrm{n\text{-}cons}}(\k, \kP)| \gg |\hat{V}^{\mathrm{cons}}(\k, \kP)| \quad \text{for} \; \kP \approx \bm{0}.
\end{equation}

\subsubsection{Energy transfer in the vicinity of the sampling mode.}
In the shear-thinning cases, the contributions $\hat{V}(\k, \kP)$ for $\kP$ close to $\k$ are small compared to the near-origin region and display no organised structure.

In the shear-thickening cases, a fundamentally different picture emerges. The coupling displays a dipole centred on the sampling wavenumber: positive values on one side and negative values on the other, with a sharp sign change at $\kP = \k$. The magnitude of this dipole is of the same order as the near-origin region. This dipole pattern is immediately reminiscent of the convective spectral energy transfer, where the nonlinear advection term redistributes energy between neighbouring wavenumbers through a similar antisymmetric structure \citep{couteauInfluenceEnergycontainingScales2026}. The resemblance strongly suggests that, in shear-thickening fluids, the viscous term actively transfers energy between nearby scales, a role traditionally attributed exclusively to the convective nonlinearity.

This dipole structure is well-defined for both $n = 1.5$ and $n = 3.0$, and persists across all sampling wavenumbers, although its absolute magnitude naturally decreases at higher $\kappa$ as the energy content of the sampling mode diminishes. We therefore state
\begin{equation}
    |\hat{V}^{\mathrm{n\text{-}cons}}(\k, \kP)| \ll |\hat{V}^{\mathrm{cons}}(\k, \kP)| \quad \text{for} \; \kP \approx \k.
\end{equation}

We note that the transfer-like dipole is observed in the shear-thickening cases but not in the shear-thinning ones. Why the conservative part becomes significant only in the shear-thickening regime remains an open question.

\subsection{Conservative symmetry and viscous energy transfer}\label{sec:symmetry}

The identification of a transfer-like dipole in the viscous coupling calls for a more rigorous test of its conservative character. As discussed in \S\ref{sec:interpretation}, the antisymmetry $\hat{X}(\k, \kP) = -\hat{X}(\kP, \k)$ is the defining property of a conservative energy transfer: it ensures that whatever energy one mode gains is exactly compensated by an equal loss in another. For the convective term, this property is satisfied exactly and is known as the detailed conservation property \citep{onsagerStatisticalHydrodynamics1949,eyinkOnsagerTheoryHydrodynamic2006}. For the viscous coupling, we have established that it does not hold in general (\ref{eq:Vfull}, \ref{eq:Vsym}). The question is therefore whether it holds approximately in the spectral region where the dipole structure is observed.

\begin{figure}
    \centering
    \includegraphics[]{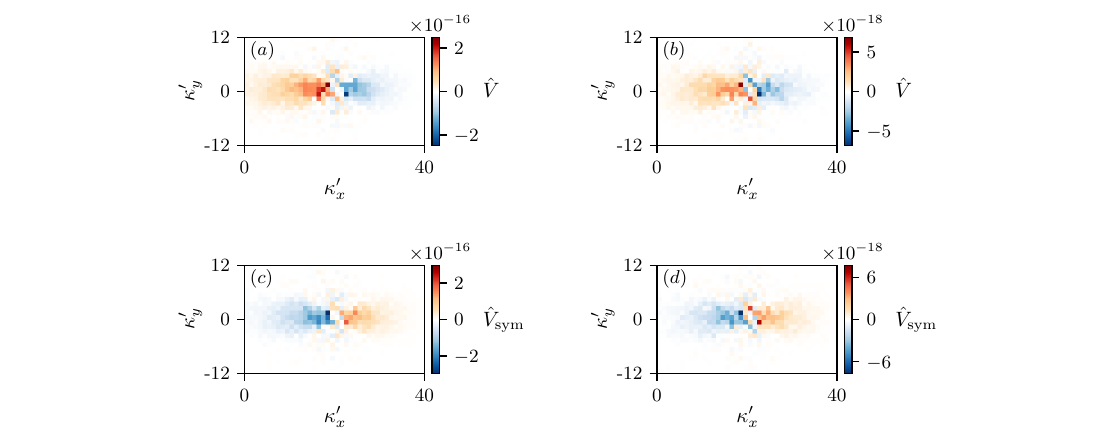}
    \caption{Viscous spectral coupling $\hat{V}(\k, \kP)$ (top) and its symmetric counterpart $\hat{V}(\kP, \k)$ (bottom) for a shear-thickening fluid with $n = 1.5$, at sampling wavenumbers $\k = [100, 0]$ (left) and $\k = [150, 0]$ (right).}
    \label{fig:symmetry_n15}
\end{figure}

\begin{figure}
    \centering
    \includegraphics[]{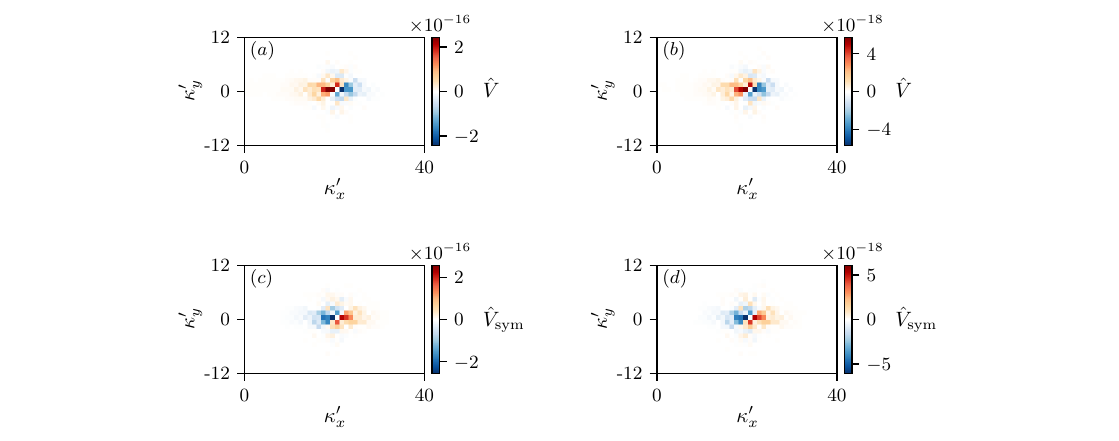}
    \caption{Viscous spectral coupling $\hat{V}(\k, \kP)$ (top) and its symmetric counterpart $\hat{V}(\kP, \k)$ (bottom) for a shear-thickening fluid with $n = 3.0$, at sampling wavenumbers $\k = [100, 0]$ (left) and $\k = [150, 0]$ (right).}
    \label{fig:symmetry_n30}
\end{figure}

Figures \ref{fig:symmetry_n15} and \ref{fig:symmetry_n30} present a direct comparison of $\hat{V}(\k, \kP)$ and $\hat{V}(\kP, \k)$, computed independently from (\ref{eq:Vfull}) and (\ref{eq:Vsym}), for $n = 1.5$ and $n = 3.0$, respectively, at two sampling wavenumbers in the dissipation range. In the vicinity of the sampling mode, both the sign and the intensity of the two quantities clearly satisfy
\begin{equation}
    \hat{V}(\k, \kP) \approx - \hat{V}(\kP, \k),
    \label{eq:approxsymmetry}
\end{equation}
in both cases. The antisymmetry is already well-defined for $n = 1.5$ (figure \ref{fig:symmetry_n15}), where the dipole is weaker, and becomes sharper for $n = 3.0$ (figure \ref{fig:symmetry_n30}), where the dipole is more intense. Near the origin, where we have claimed that non-conservative coupling dominates, no such symmetry is expected nor observed.

This result constitutes, to our knowledge, the first demonstration that energy transfer across velocity scales can arise from viscous effects in a turbulent flow. The nonlinear viscous term in the variable viscosity Navier-Stokes equations acts as a genuine inter-scale energy transfer mechanism, redistributing kinetic energy across wavenumbers without net production or destruction, in a manner analogous to the convective energy cascade. The broader implication is that the convective nonlinearity is not unique in its ability to transfer energy across scales: any nonlinear term in the momentum equation that produces a convolution product in spectral space can, in principle, carry conservative contributions. Whether these contributions are significant depends on the specific physics and must be assessed by direct computation, as we have done here.

\subsection{Mode-to-shell transfer functions}\label{sec:m2s}

\begin{figure}
    \centering
    \includegraphics[]{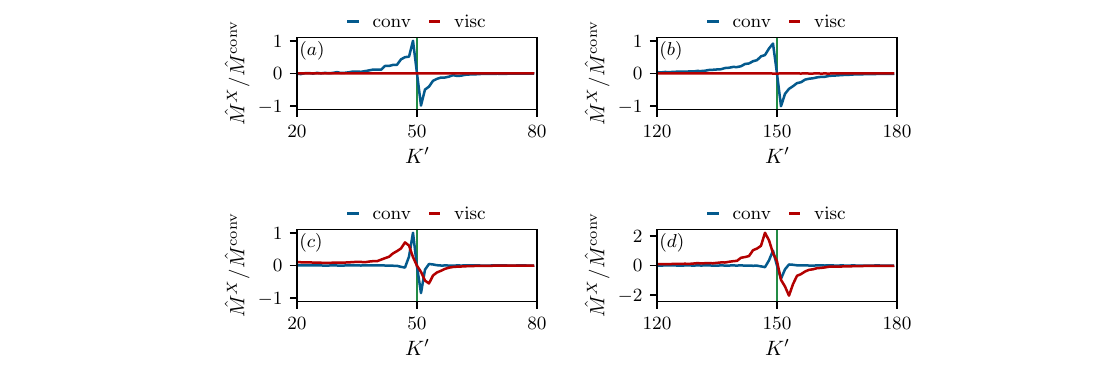}
    \caption{\textit{Mode-to-shell} transfer functions for the convective (blue) and viscous (red) spectral coupling terms. Both are normalised by the maximum of the convective transfer function. Top row: shear-thinning ($n = 0.7$), bottom row: shear-thickening ($n = 3.0$). Left column: $\k = [50, 0]$, right column: $\k = [150, 0]$.}
    \label{fig:M2Scomp}
\end{figure}

Figure \ref{fig:M2Scomp} shows the \mts transfer functions for the convective and viscous terms in a shear-thinning ($n = 0.7$) and a shear-thickening ($n = 3.0$) fluid, at two sampling wavenumbers. Note that we present $-\hat{M}^{\mathrm{conv}}$ so that positive values indicate energy received and negative values energy donated, matching the sign convention of $\hat{M}^{\mathrm{visc}}$ on the right-hand side of (\ref{eq:modalEnergyConservation}).

In the shear-thinning case, the viscous transfer function is negligible compared to the convective one at all shells and for both sampling wavenumbers. The convective transfer still displays the classical forward cascade pattern, with positive contributions from shells with $K' < \kappa$ and negative contributions from $K' > \kappa$. The spectral budget is therefore qualitatively unchanged from the Newtonian case: the convective term alone drives the energy redistribution across scales.

In the shear-thickening case, the picture changes with wavenumber. At $\k = [50, 0]$ (figure \ref{fig:M2Scomp}$c$), the viscous transfer function is already visible but remains smaller than the convective one. At $\k = [150, 0]$ (figure \ref{fig:M2Scomp}$d$), deep in the dissipation range, the viscous transfer has grown to dominate the convective transfer. This gradual takeover illustrates a key finding: the viscous energy transfer is negligible at low to intermediate wavenumbers and becomes significant only in the dissipation range, where it progressively replaces the convective transfer as the primary mechanism for inter-scale energy redistribution.

Crucially, the viscous transfer function exhibits the same forward cascade structure as the convective one: energy is received from larger scales and donated to smaller scales. The viscous term therefore participates in the forward energy cascade alongside the convective transfer.

The viscous transfer function is more spread in wavenumber space than its convective counterpart. This can be understood from the structure of the \mtm couplings. As discussed by \citet{couteauInfluenceEnergycontainingScales2026}, the convective \mtm transfer $\hat{T}(\k, \kP)$ is constrained to vanish when $\k$ and $\kP$ are aligned, due to the divergence-free condition entering through the factor $\kappa_{l} \hat{u}_{l}(\kmkP)$. This creates ``low-effectiveness'' conical regions in the $\kP$ domain. In the viscous \mtm coupling (\ref{eq:Vfull}), the factor $\kappa_{l} \hat{u}_{l}(\kmkP)$ appears as well, but is supplemented by $\kappa_{l} (\kmkP)_{l} \hat{u}_{j}(\kmkP) \hat{u}_{j}^{*}(\k)$, arising from the velocity gradient transpose in $S_{ij}$. The latter term has no geometric constraint when $\k$ and $\kP$ are aligned. Therefore, there is no global geometric suppression of the viscous transfer in the aligned directions, resulting in a broader transfer function.

\subsection{Viscous energy transfer and algebraic dissipation scaling}\label{sec:connection}

The results presented above establish a clear connection between the viscous spectral transfer and the nature of the dissipation range. In the shear-thinning cases, the viscous transfer is negligible and the exponential spectral decay is preserved, as in the Newtonian case. In the shear-thickening cases, significant viscous energy transfer emerges in the dissipation range, participating in the forward cascade alongside the convective transfer, and the classical exponential cutoff is replaced by an algebraic decay $\hat{E}(\kappa) \sim \kappa^{-\alpha}$.

\begin{table}
    \centering
    \caption{Fitted dissipation scaling range exponents $\hat{E}(\kappa) \sim \kappa^{-\alpha}$ for shear-thickening Carreau fluids.}
    \label{tab:exponents}
    \begin{tabular}{@{}ccc@{}}
    \toprule
    $n$ & $\nu_{0}$ & $\alpha$ \\
    \midrule
    1.5 & $5 \times 10^{-3}$ & 9.1 \\
    3.0 & $5 \times 10^{-3}$ & 7.9 \\
    3.0 & $1 \times 10^{-4}$ & 7.0 \\
    3.0 & $2 \times 10^{-5}$ & 7.0 \\
    \bottomrule
    \end{tabular}
\end{table}

Table \ref{tab:exponents} summarises the fitted exponents. The connection can be understood through the lens of classical scaling theory. In the Kolmogorov framework, the $-5/3$ inertial range is precisely the spectral region where the convective energy transfer is the only relevant phenomenon: the convective spectral flux is constant, meaning that kinetic energy is a conserved quantity being transferred across scales at a constant rate. The inertial range scaling is therefore a consequence of a conserved flux.

In the Newtonian dissipation range, the spectral balance reduces to a competition between the convective flux and the scale-invariant dissipation $\nu \kappa^{2} \hat{E}(\kappa)$. The constant viscosity inevitably produces a spectrum decaying faster than any power law, as the exponential cutoff is a direct consequence of scale-invariant dissipation \citep{paoStructureTurbulentVelocity1965}. In the shear-thickening case, the viscous forward cascade introduces an additional spectral flux in the dissipation range. The emergence of a power-law scaling is consistent with this picture: a power-law range indicates a regime in which there is again a conserved quantity being transferred self-similarly across scales, this time by the combined action of both convective and viscous fluxes. The algebraic decay replaces the exponential one precisely because the viscous flux sustains energy transfer to high wavenumbers, preventing the runaway dissipation that would otherwise produce the exponential cutoff. Any spectral flux in the dissipation range, whether from a physical nonlinear viscous term as here or from a modelled eddy viscosity, can in principle replace exponential cutoffs with power-law behaviour. The fact that the exponent depends on both $n$ and the Reynolds number reflects the non-universality of this dissipation scaling range, since the relative magnitudes of the convective and viscous fluxes, as well as the nonlocal structure of the dissipation itself, depend on the constitutive parameters.

\section{Discussion}\label{sec:discussion}

\subsection{Energy transfer from nonlinear terms}\label{sec:general_principle}

The central finding of this work is that the nonlinear viscous term can transfer energy across scales. It is worth stepping back to appreciate what this means in a broader context.

In classical turbulence theory, inter-scale energy transfer is attributed exclusively to the convective nonlinearity. This is because the convective term is the only nonlinear term in the constant viscosity Navier-Stokes equations. The moment a second nonlinear term appears in the momentum equation, as happens here when the viscosity varies in space, a second convolution product arises in spectral space, and with it the possibility of a second channel for inter-scale energy transfer. The derivation in \S\ref{sec:interpretation} makes this point in full generality: any nonlinear momentum term produces a convolution, and a convolution carries, in general, both conservative and non-conservative contributions. Whether the conservative part is significant, i.e., whether genuine energy transfer occurs, cannot be determined a priori and must be assessed by direct computation of the \mtm coupling.

This perspective has implications beyond the generalised Newtonian fluids studied here. In viscoelastic turbulence, for instance, the polymer stress introduces a nonlinear coupling between the velocity and conformation tensor fields. While the elastic stress has its own conservation equation and the physics are qualitatively different from the purely inelastic case \citep{valenteEffectViscoelasticityTurbulent2014}, the spectral structure of the polymer-velocity coupling must also take the form of a convolution product, and the question of whether it carries conservative contributions is, in principle, the same one we have addressed here for the viscous term. The same reasoning applies to any other nonlinear body force, stress, or source term that may appear in the momentum equation in more complex flow configurations.

We emphasise that computing $\hat{V}(\k, \kP)$ is not merely a diagnostic exercise. It is the only way to determine whether a given nonlinear term transfers energy or merely dissipates it, because the entanglement of conservative and non-conservative effects in the convolution product (see figure \ref{fig:transformVSSpectrumViscousVariable}) prevents any separation based on the physical equations alone. The \mtm coupling computation provides the empirical test: if the antisymmetry $\hat{V}(\k, \kP) \approx -\hat{V}(\kP, \k)$ is observed, transfer is occurring.

\subsection{Connection to large-eddy simulation}\label{sec:LES}

A concrete application of these findings concerns large-eddy simulation. The Smagorinsky-Lilly closure \citep{smagorinskyGENERALCIRCULATIONEXPERIMENTS1963,lillyRepresentationSmallScaleTurbulence1967} expresses the eddy viscosity as $\nu_{t} = (C_{s}\Delta)^{2}\dot{\gamma}$, which is precisely a generalised Newtonian fluid following a power law with $n = 2$. \citet{muschinskiSimilarityTheoryLocally1996} formalised this analogy by introducing the concept of the ``LES fluid'', a non-Newtonian medium whose equations of motion are those of a Smagorinsky-type LES.

The results of this work lead to a specific, falsifiable prediction: in any flow where the Smagorinsky eddy viscosity dominates over the molecular viscosity (i.e., $\nu_{t}/\nu \gg 1$), the nonlinear eddy-viscosity term should produce a viscous spectral flux in the dissipation range of the LES, and the resolved energy spectrum should exhibit an algebraic rather than exponential decay beyond the effective inertial range. In typical LES applications, the eddy viscosity can reach values two orders of magnitude larger than the molecular viscosity \citep{pope2000turbulent}, comparable to the viscosity ratios in our $n = 3.0$ cases. This prediction could be tested by performing a spectral budget analysis in a well-resolved LES of a flow with known spectral properties, such as forced homogeneous isotropic turbulence at moderate Reynolds number, and checking whether the eddy-viscosity term contributes a forward spectral flux beyond the grid cutoff.

\section{Conclusions}\label{sec:conclusions}

We have investigated the spectral energy dynamics of turbulent flows with variable viscosity using DNS of homogeneous isotropic turbulence of generalised Newtonian fluids described by the Carreau constitutive model, covering both shear-thinning and shear-thickening regimes.

The spectral evolution equations for the variable viscosity Navier-Stokes system show that the viscous term becomes nonlinear and gives rise to a convolution product in spectral space, formally analogous to that of the convective term. This convolution carries both conservative (transfer) and non-conservative (dissipation) contributions, entangled in a way that prevents closed-form separation. More generally, any nonlinear term in the momentum equation produces such a convolution, and the question of whether it transfers energy across scales can only be answered by direct computation of the \mtm coupling.

We have presented such computations for the viscous term, revealing two distinct spectral regions in the \mtm coupling $\hat{V}(\k, \kP)$: a sign-definite non-conservative region near $\kP \approx \bm{0}$, corresponding to large-scale viscosity modulation of modal dissipation, and a transfer-like dipole near $\kP \approx \k$ in shear-thickening fluids. The dipole satisfies the approximate antisymmetry $\hat{V}(\k, \kP) \approx -\hat{V}(\kP, \k)$, which is the defining signature of a conservative energy transfer. This demonstrates that inter-scale energy transfer is not the exclusive province of the convective nonlinearity: the viscous term can play the same role when the viscosity varies in space. The dipole is observed in both shear-thickening cases studied ($n = 1.5$ and $n = 3.0$) but not in the shear-thinning cases; the origin of this asymmetry remains an open question.

The \mts transfer functions confirm that the viscous transfer participates in the forward energy cascade alongside the convective transfer, growing from negligible at intermediate wavenumbers to dominant in the dissipation range. Its presence is connected to the emergence of power-law spectral decay $\hat{E}(\kappa) \sim \kappa^{-\alpha}$ in shear-thickening fluids, replacing the classical exponential cutoff. In shear-thinning fluids, where no viscous transfer is observed, the exponential decay is preserved. The exponent $\alpha$ depends on both the power index $n$ and the Reynolds number, with values converging towards approximately $7$ at higher Reynolds numbers for $n = 3.0$.

The turbulence statistics confirm that Kolmogorov scaling extends to both shear-thinning and shear-thickening fluids upon rescaling with the mean viscosity, generalising to the shear-thickening regime a result previously reported for shear-thinning fluids.

The formal equivalence between the Smagorinsky LES closure and a shear-thickening power-law fluid with $n = 2$ leads to a concrete prediction: LES with eddy-viscosity models operating at $\nu_{t}/\nu \gg 1$ should exhibit viscous spectral transfer and algebraic decay in the resolved dissipation range, a prediction that can be tested by spectral budget analysis.


\vspace{0.5cm}\noindent\textbf{Acknowledgments:} The authors are grateful to J.A. Domaradzki for insightful discussions throughout the course of this study.


\vspace{0.2cm}\noindent\textbf{Funding:} The authors gratefully acknowledge the financial support by the Swiss Federal Office of Energy (SFOE) in the project HPH2HD, contract number S1/502196\-01.

\vspace{0.2cm}\noindent\textbf{Use of AI:} The authors acknowledge the use of generative AI tools (Anthropic's Claude) to assist in drafting and refining portions of the manuscript. All scientific content, analysis, and conclusions are the sole responsibility of the authors, who have reviewed and edited the final text.


\clearpage
\appendix
\setcounter{figure}{0}

\section{Kinetic energy spectra}
\label{sec:app:KineticEnergySpectra}

\begin{figure}[!h]
    \centering
    \includegraphics[]{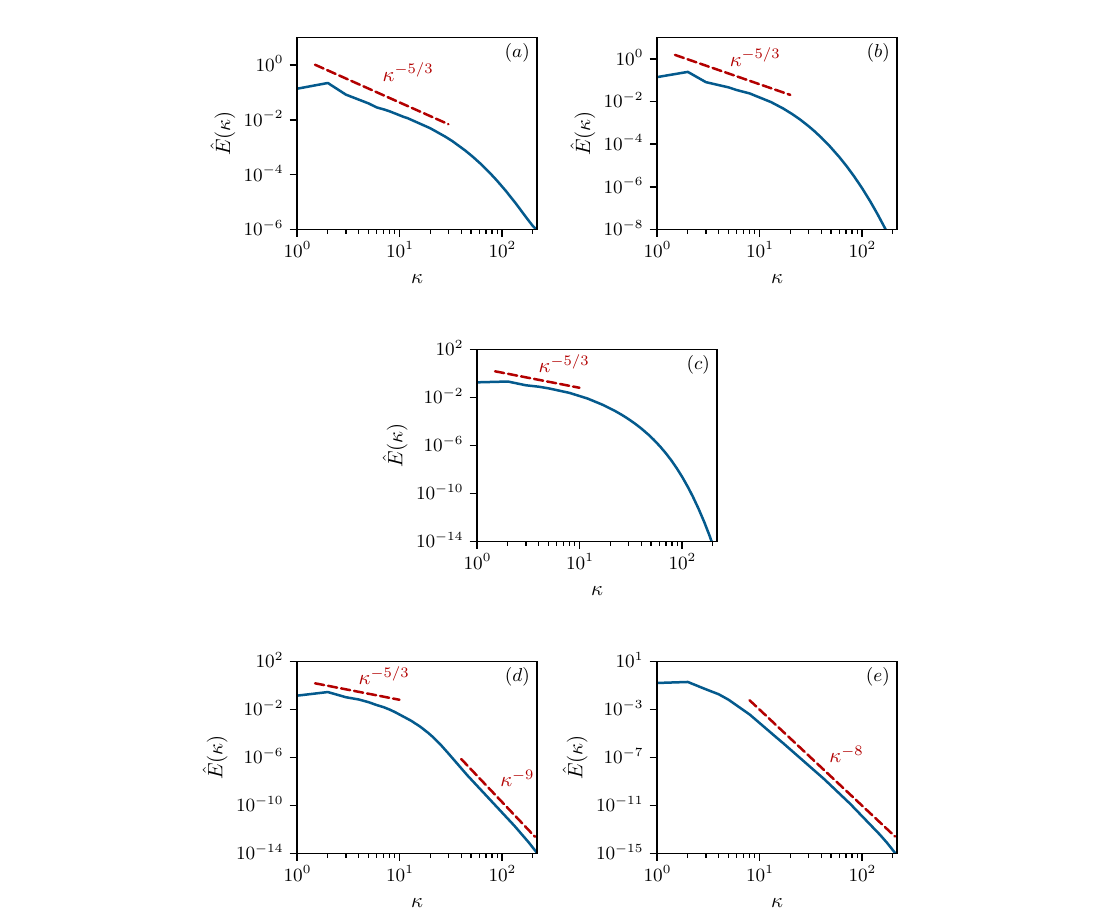}
    \caption{Kinetic energy spectra $\hat{E}(\kappa)$ with scaling ranges highlighted.}
    \label{fig:app:kinEnergySpectra}
\end{figure}


\clearpage
\bibliographystyle{unsrtnat}
\bibliography{references.bib}

\end{document}